\documentclass[english,a4paper,11pt]{article}
\usepackage[margin=3cm]{geometry}
\usepackage[latin1]{inputenc}
\usepackage{babel}
\usepackage{bm}
\usepackage{amsmath,amsthm}
\usepackage{latexsym}
\usepackage{booktabs}
\usepackage[final]{graphicx}
\DeclareGraphicsExtensions{.jpg,.jpeg,.pdf,.png,.mps}
\usepackage{epsfig}
\usepackage[round]{natbib}
\usepackage{natbib}
\setcitestyle{aysep={}} 
\usepackage{rotating}
\usepackage{color}
\usepackage{xcolor}
\usepackage{colortbl}
\usepackage[bitstream-charter]{mathdesign}
\usepackage[T1]{fontenc}
\usepackage{threeparttable}
\usepackage{float}
\usepackage{url} 
\usepackage{appendix}
\usepackage{hyperref}
\usepackage{pdflscape}
\usepackage[misc]{ifsym}
\hypersetup{
     colorlinks   = true,
     citecolor    = blue
}

\usepackage{silence}
\title{Accounting carbon emissions from electricity generation: a review and comparison of emission factor-based methods}

\author{$\mathrm{Marina \ Bertolini}^\mathrm{*},  
	\  \mathrm{Pierdomenico\ Duttilo}^\mathrm{*,\hspace{0.5mm}\textrm{\Letter}},
	\  \mathrm{Francesco\ Lisi}^\mathrm{*}$\\  
	$^\mathrm{*}$\small{\emph{Department  of Statistical Sciences,  University of Padua, Italy}}\\
    $^\mathrm{\textrm{\Letter}}$\small{\emph{Corresponding author: \href{mailto:pierdomenico.duttilo@unipd.it}{\textcolor{blue}{pierdomenico.duttilo@unipd.it}}}}
}
\date{}

\begin{document}
	\maketitle
\begin{abstract}
	
Accurate estimation of greenhouse gas (GHG) is essential to meet carbon neutrality targets, particularly through the calculation of direct CO$_2$ emissions from electricity generation. This work reviews and compares emission factor-based methods for accounting direct carbon emissions from electricity generation. The emission factor approach is commonly worldwide used. Empirical comparisons are based on emission factors computed using data from the Italian electricity market. The analyses reveal significant differences in the CO$_2$ estimates according to different methods. This, in turn, highlights the need to select an appropriate method for reliable emissions, which could support effective regulatory compliance and informed policy-making. As concerns, in particular, the market zones of the Italian electricity market, the results underscore the importance of tailoring emission factors to accurately capture regional fuel variations.
	
\noindent 
\hspace{1cm}\\
\emph{Keywords}: carbon emission, electricity generation, direct carbon emission, emission factor-based methods. 
\end{abstract}

\section{Introduction}\label{sec.Introduction}
Estimation of greenhouse gas emissions (GHG, usually expressed as tons of equivalent CO$_2$) is a particularly relevant activity in the context of the green energy transition and, more generally, to establish the drivers of sustainability policies \citep{co2_equivalent}.
Carbon emissions are deeply related to climate change \citep{ipcc2021} and pollution issues in several sectors \citep{LI2017976,LEROUTIER2022105941,FREITAG2021}. Therefore, they can be seen as relevant metrics to measure progress or ``steps back'" with respect to institutional policy objectives year by year. It is important to note that the initial international targets for managing climate change were directly related to emission levels, using 1990 as the base year, as outlined in the Kyoto Protocol \citep{kyoto1997}. As the loyalty to mitigate climate change and reduce emissions continues, other international agreements have renewed the commitment to pursue this goal \citep{paris2015} setting the objective of carbon neutrality by 2050. In addition, these targets align with the sustainable development goal 7 (SDG 7), which aims to ensure access to affordable, reliable, and sustainable energy for all people, promoting sustainable practices in the energy sector, which are essential to reduce GHG emissions and contrast climate change \citep{SDG7}.
\subsection{Literature Review}\label{subsec.Literature}
Over the years, despite the increasing penetration rate of renewable production sources, traditional power plants have continued to give their contribute to the electrical system and meet energy demand \citep{busch2023,HASSAN2023,KHALID2024}. For several years, their role will remain crucial due to the need for stability of electricity systems and their ability to provide additional services (e.g., the capacity market). Thus, it is essential to understand and quantify the impact of power plants in terms of emissions.\\
Furthermore, since the electricity production sector is crucial in the energy transition process, it is included in the European Emission Trading System (ETS). The industrial sectors included in the ETS system have been calculating their emissions for years, using different methods, often paying more attention to the need to reach an estimation than to the accuracy of the calculation, also to avoid excessive burden for companies \citep{eu_ets,SHOBANDE2024}. Even if the calculated emissions are then verified by an independent third party, a deeper understanding of the relevance of different accounting methods in determining final results is suitable.
The GHG Protocol, developed by \cite{ghgprotocol}, is the principal global framework for carbon accounting and the most widely recognised international standard \citep{LI2024122681}. To differentiate between direct and indirect emission sources, improve transparency, and provide a comprehensive framework that supports various organisations and climate-related strategies, three scopes have been defined for GHG accounting \citep{Ranganathan2004}.

Scope 1 includes direct emissions occurring within a specified accounting boundary, such as those from the combustion of fuels used to generate electricity. Scope 2 ``includes indirect energy-related emissions that occur outside the accounting boundary" \citep{LI2024122681} such as, for example, emissions resulting from electricity consumption by users. Finally, Scope 3 incorporates all other indirect emissions not captured by Scopes 1 and 2.

Tracking Scope 3 emissions presents significant challenges, as it requires monitoring a wide range of emissions across the industrial supply chain \citep{shan2018rapid, WEI2020119789}. Depending on the analytical approach and the definition of ``boundaries", a carbon emissions inventory can be compiled using these three scopes \citep{Wiedmann2021}.

Carbon emissions accounting can be a complex task due to several factors, including comparability (both between different entities and for the same entity over time), fairness, and the potential for greenwashing, which must be identified to ensure market transparency \citep{Treepong2024}. As emission accounting forms the foundation for expanding markets, it is subject to legal obligations and limits that apply to various categories of stakeholders, alongside national policy orientations. Thus, emission accounting methods and their differences are important topics of discussion.

A recent debate has arisen about consumption-based accounting (CBA) versus production-based accounting (PBA) for carbon emissions \citep{fan2016exploring,FERNANDEZAMADOR2017269,Afionis2017}. 
CBA refers to the measurement of GHG emissions resulting from the consumption of all goods and services in a specified region, regardless of their origin \citep{Wiedmann2016}. In contrast, PBA focuses on emissions generated by production processes of goods and services,
within a particular geographical area regardless of where they are consumed. 

\cite{FRANZEN201834} compare production-based and consumption-based emissions, examining the reasons for any differences between the two approaches. They investigated whether carbon spillover occurs from developed to developing countries using the data from 110 countries and focusing on the differences between OECD and non-OECD members. The findings indicate that the differences between the two accounting methods are generally small for most countries.

Usually, direct carbon emissions (Scope 1) are calculated using the PBA approach, whereas indirect emissions (Scope 2) are based on the CBA principles \citep{fan2016exploring}. National greenhouse gas (GHG) inventories adhere to IPCC guidelines, applying PBA to measure emissions from domestic use of oil, coal, and gas, which includes emissions from households, industrial production, and electricity generation. In contrast, the protocols described by \cite{bsi2013} recommend the use of CBA. This dual accounting approach facilitates a more comprehensive understanding of the environmental impact associated with both the perspective of consumption and production activities \citep{fan2016exploring}.

\cite{LI2024122681} provide a comprehensive review of carbon emissions accounting in the electricity power industry, focusing on the approaches within Scope 1 and Scope 2. The authors evaluate the strengths and limitations of existing accounting methods, highlighting recent advancements that improve the precision and reliability of carbon emissions assessments.

\begin{figure}[h]
	\centering
	\includegraphics[width=1\textwidth]{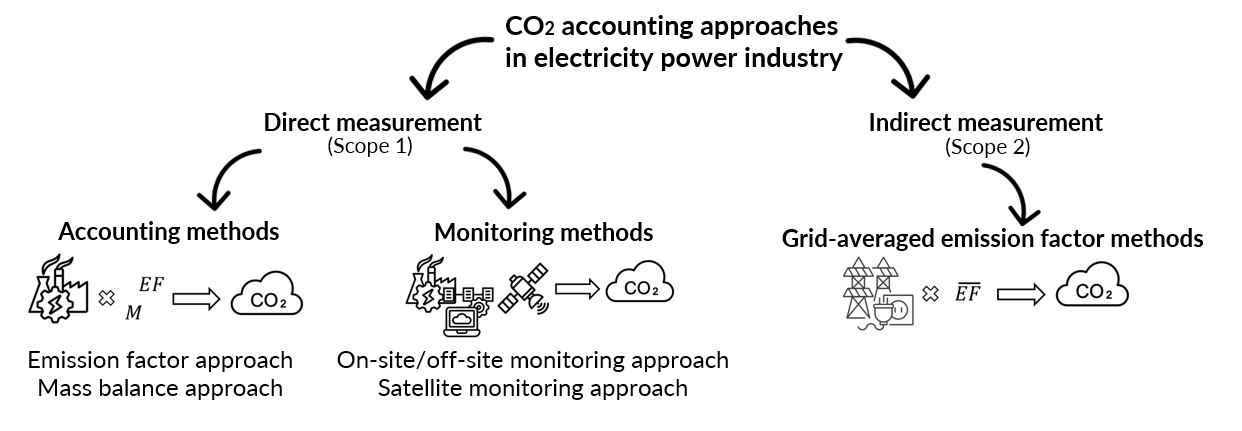}
	\caption{Classification of the main carbon emissions accounting approaches in the electricity power industry.}\label{fig.co2info}	
\end{figure}

Figure \ref{fig.co2info} provides an infographic classification of the main carbon emissions accounting approaches in the electricity power industry.
\\
The direct measurement approach is typically divided into accounting and monitoring methods \citep{LI2024122681}. The emission factor approach calculates carbon emissions based on fuel combustion, directly related to electricity generation, while the mass balance approach, grounded in the principle of mass conservation, ensures that carbon inputs and outputs remain consistent in the combustion process. Monitoring methods include on-site systems such as continuous emission monitoring systems (CEMS) and off-site techniques \citep{Fan2009,Zhang2024}. Additionally, satellite-based approaches track atmospheric carbon levels and temporal-spatial trends using sensors such as Fourier transform infrared spectrometers and laser radars \citep{XIA201643,rs16183394}. 

Several scholars have proposed grid-averaged emission factor methods with various boundary definitions to assess indirect carbon emissions from power systems \citep{SOIMAKALLIO201213,Miller_2022,GAO2023137618,LI2024122681}. These include the standard grid-averaged emission factor method and some variants that incorporate time-varying dynamics (the time-varying grid-averaged emission factor) and electricity trading behaviour (the modified grid-averaged emission factor accounting for trading patterns).

The problem of different possible estimates of carbon emissions was already discussed by \cite{LEE2018247} with reference to waste management, while \cite{zubair2023systematic} suggested a systematic review of the methods used to calculate carbon emissions in the transportation sector. 

\subsection{Paper contributions}\label{subsec.contributions}
Carbon emissions are of great relevance in current and future economies, potentially having a role in determining the competitiveness of firms \citep{MENG2018,GRIFFIN2018,RISSMAN2020}. In recent years, carbon emissions accounting has become increasingly relevant, and in the future it will further increase its role because it affects the reputation of companies, becoming increasingly called to present themselves as ``sustainable'' \citep{ahmad2024}. 

The present work investigates various methodologies to estimate CO$_2$ emissions from electricity generation with a focus on accounting methods based on the emission factor approach. Although this is the most widely used accounting method \citep{Hiete2001}, to the best of our knowledge, a review of the latest emission factor-based methods or a quantitative comparison between methods are still lacking.

Starting with the foundational methods provided by the Intergovernmental Panel on Climate Change (IPCC), a comprehensive literature review of current academic studies using emission factor-based methods is conducted in order to understand their evolution and application in the context of electricity generation.

In addition, we perform a comparative analysis of the most widely used emission factor-based methods using zonal data from the Italian electricity market. Our goal is to quantify the differences between methods. Furthermore, a detailed analysis of the evolution of CO$_2$ emissions in Italy over the past few years is provided.\\
Our findings highlight significant differences in precision between methods and underscore the importance of adopting accurate accounting methods for effective policy-making and regulatory compliance.

The rest of the paper is organised as follows. Section \ref{sec.IPCC} described the IPCC methodology for the calculation of CO$_2$ emissions; Section \ref{sec.litterature} recalls the application of the IPCC Tiers in the literature. A comparison among different estimation techniques is presented in Section \ref{sec.comparison}. Section \ref{sec.policy} highlights policy recommendations based on results to support emission reduction. Finally, Section \ref{sec.conclusions} concludes by highlighting the key contributions of the study and emphasising its relevance.

\section{The emission factor approach}\label{sec.IPCC}
Estimating CO$_2$ emissions from electricity generation is based on several key components. First, the activity data represent either electricity generation or fuel consumption. Electricity generation refers to the total amount of electricity produced by a specific power plant or within a market zone \citep{IPCC2006}. It is typically measured in kilowatt hour (kWh) or megawatt hour (MWh). Fuel consumption denotes the total amount of fuel used by the electricity generation unit and is usually measured in mass units such as kilogrammes (kg) or in energy units such as terajoule (TJ).
 
The emission factor is another essential element, indicating the quantity of CO$_2$ emissions generated per unit of electricity produced or fuel consumed. It is expressed in tons of CO$_2$ per megawatt-hour (tCO$_2$/MWh) or tons of CO$_2$ per terajoule (tCO$_2$/TJ). Emission factors vary according to several characteristics, such as fuel type and quality, because their different chemical properties and carbon content affect pollutant emissions \citep{IPCC2006}. 

\cite{Hiete2001} provide a review of the emission factors of fossil fuels. Coal emissions vary by type: hard coal, with a higher carbon content, releases more CO$_2$ than lignite or brown coal, which contains more moisture and a lower energy density. Crude oil has a more consistent emission factor, as the carbon content varies in the range 83-87\%. However, the specific gravity of the crude oil, often measured using API (American Petroleum Institute) gravity, can cause slight variations. Lighter crude oils, which contain more hydrogen and less carbon, tend to release slightly less carbon emissions than heavier crude oils. Natural gas emissions also depend on composition; dry gas, mainly methane, emits less CO$_2$ than wet gas, including heavier hydrocarbons such as propane and butane. Additional hydrocarbons and impurities, such as CO$_2$ and nitrogen, increase the emissions in the wet and associated gas.

Combustion temperature also plays a role, as higher temperatures typically result in more complete combustion, reducing emissions of carbon monoxide. In contrast, incomplete combustion can lead to an increase in unburned hydrocarbons and particulates \citep{KAKAEE201464,maurya2017}. 

Technology-specific factors significantly affect emissions. They include the efficiency of power plants, the combustion technology in use, and the presence of pollution control systems such as scrubbers and catalytic converters that reduce emissions by capturing pollutants before they are released into the atmosphere \citep{moazzem2012,ZHAO2015}. As a consequence, emission factors can vary widely between power plants, regions, and countries using different fuel types, combustion conditions, and technology. When specific emission factors are not readily available, the default values proposed by international organisations can be applied \citep{IPCC2006}. 

In mixed fuel generation systems, the energy share, i.e. the proportion of electricity produced by each fuel type, is essential for calculating weighted average emissions when multiple fuels are used \citep{SUMABAT2016}. Another key parameter is the carbon oxidation fraction, which represents the percentage of carbon in the fuel that is converted to CO$_2$ during combustion. This percentage is typically close to 100\%, and only a small percentage of carbon remains unburned \citep{IPCC2006}.

The net calorific value is a crucial parameter to understand how much energy can actually be produced by fuel \citep{etcacc2003,IPCC2006} and, thus, to convert fuel consumption into energy terms. It reflects the energy available after subtracting the latent heat lost during combustion and is usually measured in energy per unit of fuel, i.e. megajoules per kilogramme (MJ/kg). 

Emissions are also affected by the molecular weight ratio of CO$_2$ to carbon, which is approximately 3.67 (44/12) and is used to convert the carbon mass emitted into the corresponding mass of CO$_2$ \citep{Hiete2001}.

The Intergovernmental Panel on Climate Change (IPCC) provides authoritative assessments on climate change science and its potential impacts, as well as strategies to mitigate climate change. The \cite{IPCC2006} guidelines also define the methodological framework to account GHG emissions in various sectors. A straightforward methodological approach to measure GHG emissions ($E$) involves combining information on the extent of human activities referred to as activity data ($AD$) with coefficients that quantify emissions per unit of activity ($EF$)

\begin{equation}\label{eq.baseipcc}
	\textit{E}=\textit{AD}\times\textit{EF}.
\end{equation}

In Eq. (\ref{eq.baseipcc}), the specification of \textit{AD} and \textit{EF} depends on the type of GHG emissions, the sector under consideration, and the available data. In calculating emissions from stationary sources, the sectoral approach usually multiplies the fuel consumption by the corresponding emission factor. In countries with advanced accounting systems, the discrepancy in energy activity data is typically kept within 5\%, while in countries with less developed systems it can reach 10\% \citep{IPCC2006,LI2024122681}.

\begin{figure}[H]
\centering
\includegraphics[width=1\textwidth]{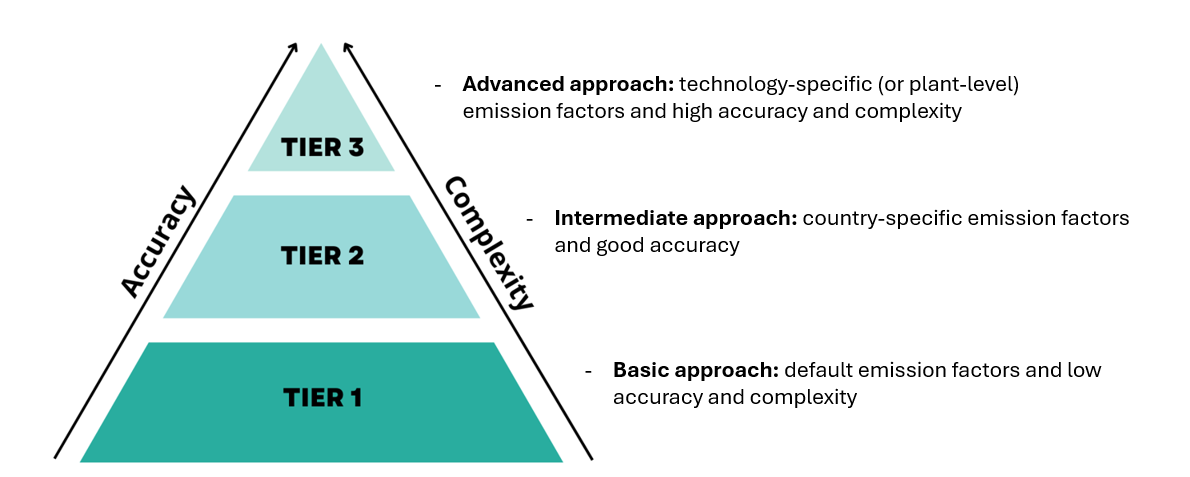}
\caption{IPCC Tier methodology - assessing CO$_2$ estimates.}\label{fig.Tiers}	
\end{figure}

The \cite{IPCC2006} guidelines also allow for more complex approaches which can be divided into three different Tiers. In Figure \ref{fig.Tiers}, the IPCC framework is sketched with a three-level pyramid with increasing complexity and accuracy, where ascending from bottom to top, the complexity of the data increases together with the accuracy of the estimates. Tiers 2 and 3, often called higher-tier methods, are more accurate than Tier 1 because they require country-specific or technology-specific information. In contrast, Tier 1 yields the least accurate estimates because it relies on default emission factors, which do not account for specific conditions at the plant or country level. Although one might consider directly adopting Tier 3, the required data for its calculation are not always available. Consequently, the selection of the Tier has to be based on both the quantity and quality of the available data.

This three-tier approach can also be applied to estimate CO$_2$ emissions from electricity generation, where Tier 1 uses default emission factors for a basic estimate, Tier 2 improves accuracy with country-specific factors, and Tier 3 provides the most detailed estimates by incorporating the specific characteristics of thermal power plants. The different tiers will be illustrated in detail in the following.

Coherently with Eq. (\ref{eq.baseipcc}) the Tier 1 provides a basic estimation method for estimating CO$_2$ emissions by using default emission factors
\begin{equation}
\label{eq.tier1}
	E_{t,f}=G_{t,f} \times EF_f,
\end{equation}
where $E_{t,f}$ are CO$_2$ emissions by fuel type $f$ ($f=1,...,F$) at time $t$, $G_{t,f}$ is the amount of fuel $f$ combusted to generate electricity at time $t$ and $EF_f$ is the default emission factor.
The latter is typically provided by international institutions such as the IPCC.
Likewise, $G_{t,f}$  denotes the amount of electricity produced from the type of fuel $f$ at time $t$ \citep{ANG2011,SUMABAT2016,EBERLE2020}.

The total CO$_2$ emissions at time $t$ are then calculated as the sum of CO$_2$ emissions over fuel type $f$
\begin{equation}
	E_{t}=\sum_{f=1}^F E_{t,f}.
\end{equation}

This expression gives a basic estimate of CO$_2$ emissions without accounting for country-specific information.

The Tier 2 offers improved accuracy by including country-specific emission factors, $EF_{f,i}$, accounts for local fuel properties and other country-specific information $i$, $(i=1,...,I)$
\begin{equation}
\label{eq.tier2}
	E_{t,f}=\sum_{i=1}^I G_{t,f,i}  \times EF_{f,i},
\end{equation}
A country-specific emission factor might be the same as the default one or different. In general, as it considers more specific information, it is expected to be more accurate, with a smaller range of uncertainty compared to the default emission factor $EF_f$. In this context, \cite{IPCC2006} suggests, as a best practice, to compare any country-specific emission factor with default values.

The Tier 3 approach aims at the highest accuracy by considering the specific technologies and conditions under which fuel is combusted. The drawback of this approach is that it requires detailed data on the type of fuel, combustion technology, operating conditions, control technologies, and the age and maintenance of the power plant \citep{IPCC2006}. Clearly, this information is not always available. The specific technology $l$ ($l=1,...,L$) includes any device, combustion process, or fuel property that affects CO$_2$ emissions. The corresponding emissions level is given by:
\begin{equation}
	E_{t,f}=\sum_{l=1}^L G_{f,t,l} \times EF_{f,l}.
\end{equation}

Fuel consumption data are divided according to the different technologies in use and technology-specific emission factors are applied for each combination. If the direct fuel consumption data per technology is unavailable, it can be estimated as follows
\begin{equation}
	G_{t,f,l}=G_{f,t} \times P_{l},
\end{equation}
where $P_{l}$ is the fraction of the full source category engaged by a given technology, i.e. the technology penetration rate. According to \cite{IPCC2006} the term $P_{l}$ \textit{"can be determined on the basis of output data such as electricity generated which would ensure that appropriate allowance was made for differences in utilisation between technologies"}.

\section{Literature review of emission factor approaches}\label{sec.litterature}
Starting from the IPCC guidelines, various contributions in the literature have aimed to enhance CO$_2$ emissions estimates. These studies focus on different aspects, such as the research objectives, the selection of tiers, the country context, the time period, the quantities, and the measures used. The approaches discussed so far require different types and amounts of information. Specifically, complex methodologies aligned with high Tiers involve more complex calculations and detailed data collection, leading to potentially more accurate estimates. 
In this section, we will review the most emission factor-based methods found in the literature.

Table \ref{tab.contr} provides a summary of the most significant contributions in the literature on CO$_2$ emissions estimation, classified according to the IPCC Tier approach.

\begin{table}[!ht]
	\caption{Summary of relevant contributions by IPCC tier methodology.}
	\label{tab.contr}
	\centering 
	\begin{tabular}{ll}
		\toprule
		Tier & Contribution\\	
		\midrule
		Tier 1& \cite{SHRESTHA2009,MALLA2009,liu2012,ZHANG2013}\\
		&\cite{emodi2015,YANG2016,SUMABAT2016}\\
		&\cite{YAN2016,Zhao2017,Beidari2017}\\
		& \cite{MOUSAVI2017,WANG2017,WANG2018,WANG2021}\\
		Tier 2 & \cite{Tian2016,EBERLE2020}\\
		Tier 3 & \cite{GU2015,ALIPRANDI2016,MARCANTONINI2017}\\
		&\cite{Beltrami2021b,Beltrami2021a,carbonm1}\\
		\bottomrule
	\end{tabular}
\end{table}

 Most studies have used Tier 1 methods, including those by \cite{SHRESTHA2009,MALLA2009,liu2012,ZHANG2013,emodi2015,YANG2016,SUMABAT2016,YAN2016,Zhao2017,Beidari2017,MOUSAVI2017,WANG2017,WANG2018,WANG2021}. Tier 2 approaches are less common, as seen in \cite{Tian2016,EBERLE2020}, and the Tier 3 methods are used only in a limited number of studies, such as \cite{GU2015,ALIPRANDI2016,MARCANTONINI2017,Beltrami2021b,Beltrami2021a,carbonm1}. The prevalence of Tier 1 approaches highlights the effort and the complexity of using methodologies, such as Tier 2 and Tier 3, which require more granular and detailed data.

Furthermore, a significant part of the literature focusses on the decomposition of CO$_2$ emissions, as seen in the work of \cite{SHRESTHA2009,MALLA2009,liu2012,ZHANG2013,emodi2015,YAN2016,Beidari2017,Zhao2017,MOUSAVI2017,WANG2017,WANG2018,WANG2021}. These studies aim to identify and analyse the key factors driving changes in carbon emissions over time using decomposition analysis techniques. Other research has focused on the reduction of marginal emission factors \citep{Beltrami2021a} and the impact of renewable energy sources on reducing CO$_2$ emissions \citep{ANG2011,ALIPRANDI2016,Beltrami2021b}. These contributions provide valuable information on the various strategies and methodologies employed to address the pressing issue of carbon emissions.

In \cite{MALLA2009}, the estimation of CO$_2$ emissions for the seven major Asia-Pacific and North American countries (Australia, Canada, China, India, Japan, South Korea and the US) and for the period 1971-2005, is based on the default emission factors \citep{IPCC2006}, according to the following formula:
\begin{equation}
\label{eq.of}
	E_{t,f}=G_{t,f} \times EF_f \times O_f,
\end{equation}
where $O_f$ is the fraction of carbon oxidized of the fuel type $f$ during combustion. The $O_f$ values used by the author are based on \cite{IEA2007}. Specifically, for coal and coal products, 98\% of carbon is oxidised, for oil and oil products it is 99\%, for natural gas it is 99.5\%, and for peat it is 99\%. Since this ratio is usually high, Tiers 1-3 assume complete oxidation of the carbon in the fuel, assigning $O_f=1$. However, this assumption may not be accurate for some solid fuels, where the actual fraction of carbon oxidised could be significant. In this context, good practise involves the use of high Tiers and country-specific values \citep{IPCC2006}. 

In estimating CO$_2$ emissions from electricity generation in China for the period 1997-2040, \cite{WANG2021} consider also the fraction of carbon oxidised by the net calorific value of the fuel type ($NCV_f$)
\begin{equation}
\label{eq.nvc}
	E_{t,f}=G_{t,f} \times EF_f \times NCV_f \times O_f.
\end{equation}
The net calorific value, as defined by \cite{IPCC2006}, reflects the heating potential of a fuel, accounting for the energy released during combustion adjusted for the latent heat of vaporisation of water. Typically, NCV is approximately 95\% of the gross calorific value for liquid fossil fuels, solid fossil fuels, and biomass, and approximately 90\% for natural gas.

\cite{liu2012,ZHANG2013,emodi2015,YAN2016,Zhao2017,WANG2017} multiply the fraction of carbon oxidized by $M$ the molecular weight of CO$_2$ divided by the atomic weight of carbon 
\begin{equation}
\label{eq.m}
	E_{t,f}=G_{t,f} \times EF_f \times O_f \times M.
\end{equation}
Usually, the ratio $M$ is fixed to $44/12=3.6667$ (molecular weight of CO$_2$/atomic weight of carbon) \citep{Hiete2001,Zhao2017}. This ratio is crucial in estimating CO$_2$ emissions because it converts the mass of carbon into the corresponding mass of CO$_2$, accurately accounting for the additional weight contributed by oxygen atoms in CO$_2$. \cite{Tian2016} calculate CO$_2$ emissions for Guangdong province (China) over the period 2005-2014 using the following formula:
\begin{equation}
	E_{t,f}=G_{t,f} \times EF_f \times O_f \times M + Q_p \times CI_p.
\end{equation}
where $Q_p$ represents the net purchased electricity of Guangdong province, while $CI_p$ is the average CO$_2$ emission factor of the southern power grid of Guangdong province. 

Incorporating the term $Q_p \times CI_p$ significantly improves the estimate of CO$_2$ emissions by accounting for imported electricity emissions. This consideration is essential, as many regions rely on electricity imported from other grids. By including this term, the calculation ensures that the total emissions attributed to the province cover both locally generated and imported electricity. This approach provides a more comprehensive and precise reflection of the province's overall carbon footprint, preventing the underestimation of emissions that would occur if only local generation were considered.

\cite{EBERLE2020} examine the impact of various sources of electricity generation on US CO$_2$ emissions by introducing scenario-specific emission factors that cover the period 2015-2040. To achieve this, CO$_2$ emissions are estimated for two exploratory scenarios $s$, low and high emissions
\begin{equation}
	\begin{split}
		&E_{t,f,s}=G_{t,f,s}\times EF_{f,s}=(C_{t,f,s}\times CF_{f,s})\times EF_{f,s},\\
	\end{split}
\end{equation}
where $C_{t,f,s}$ is the capacity of the generation source $f$, while $CF_{f,s}$ is the capacity factor of generation source $f$ (the fraction of hours in a year generating power). 

\cite{SUMABAT2016} provide estimates of CO$_2$ emissions for the Philippines in the period 1991-2014. The study incorporates $\alpha_{t,f}$, which represents the energy share of the fuel type $f$ at time $t$, indicating the proportion of total electricity generated by each fuel type. The CO$_2$ emissions are calculated using the formula:
\begin{equation}
\label{alpha}
	E_{t,f}=G_t \times \alpha_{t,f} \times EF_f.
\end{equation}
where $G_t$ is the total electricity generated at time $t$ and $\alpha_{t,f}=G_{f,t}/G_t$.

\cite{Beltrami2021b,Beltrami2021a} estimate the hourly CO$_2$ emissions at the plant-level $p$ for Italy in 2018 as follows:
\begin{equation}
\begin{split}
\label{eq.beltrami}
		&E_{t,f,p}= EF_f \times \lambda \times g_{f,p}(G_{t,p}),\\
		&g_{f,p}(G_{t,p})=\sum \alpha_{f,p} \biggr(c_{2,f,p}G^2_{t,f,p}+c_{1,f,p}G_{t,p}+c_{0,f,p}\biggr),
\end{split}
\end{equation}
where $G_{t,p}$ is the hourly accepted electricity generation \footnote{In the day ahead market.} of the power plant $p$, $\lambda \times g_{f,p}$ is the average plant-specific
technical efficiency parameter, $g_{f,p}(G_{t,p})$ is a quadratic fuel consumption function \citep{MARCANTONINI2017,creti2019}, and $\alpha_{f,p}$ is the percentage mix of the used fuels. In contrast to \cite{SUMABAT2016}, the coefficient $\alpha_f$ is specific to each plant. The quadratic fuel consumption is used to estimate the hourly fuel consumption $g_{f,p}$ by the relevant thermometric unit. It is characterised by three non-negative coefficients for the hourly consumption quadratic curves $\bigr(c_{2,f,p},c_{1,f,p},c_{0,f,p}\bigr)$ that are related to the fuels used in the production mix. \cite{MARCANTONINI2017} shares the same methodology as \cite{Beltrami2021a,Beltrami2021b}. This is a good example of a Tier 3 approach, but the sources of detailed input data, particularly for plant-level specifics, are not completely clear in the work. This raises questions about the availability and accessibility of such granular data, which is crucial for Tier 3 calculations.

According to \cite{ALIPRANDI2016} CO$_2$ emissions are calculated with the following equation, assuming a stoichiometric combustion
\begin{equation}
	E_{t,f}=\frac{G_{t,f} EF_f}{\eta\text{LHV}},
\end{equation}
where $\eta$ is the variable efficiency of the generator, and LHV is the Lower Heating Value, for which a value of 25 MJ/kg and 35 MJ/kg is taken for coal and natural gas, respectively. The authors apply this methodology to estimate CO$_2$ emissions for Italy during the period 2012-2013.

\cite{carbonm1} introduced an estimation method that corrects the default emission factor of the IPCC by incorporating country-specific baseline emissions from 2019. This reflects the unique mix of power generation technologies, fuel types, and efficiencies present in a country's energy sector. CO$_2$ emissions with the adjusted emission factor $EF_{f,adj}$ are calculated as follows:
\begin{equation}\label{eq.carbonm}
\begin{split}
&E_{t,f}=G_{t,f} \times EF_{f,adj},\\
&EF_{f,adj} = EF_f \times \frac{E_{\text{2019-baseline}}}{\sum(G_{2019,f} \times EF_f)},
\end{split}
\end{equation}
where $E_{\text{2019-baseline}}$ is the country-specific baseline emissions \citep{carbonm2,carbonm3}. For Italy, baseline emissions are provided by national institutions like \cite{terna1} or \cite{ISPRA}.

\section{{Empirical comparison among emission factor approaches}}\label{sec.comparison}

As we saw in Section 3, various methods for estimating CO$_2$ emissions have been used in the literature. However, they refer to different countries and different time periods. This makes the comparison among methods less useful and potentially unfair. In particular, it makes quite difficult to evaluate to which extent each method possibly underestimates the actual emissions level. 
Since a systematic and homogeneous comparison has not yet been provided in the literature, in this section we study and compare the performance of six different methods, among those previously described, using the same data set. They are listed in Table \ref{tab.methods}, together with their classification in terms of Tier. Table \ref{tab.fuelp}, instead, details the fuel parameters and the emission factors used to evaluate the CO$_2$ emissions with the different methods.
  
\begin{table}[h]
	\caption{Methods used for comparison. Eq. refers to the equation in the paper describing the model; Tier classifies the model with respect tiers; EF specifies the source of the emission factor.}
	\label{tab.methods}
	\centering 
	\resizebox{12cm}{!}{
		\begin{tabular}{lcccccc}
			\toprule
			& Method 1 & Method 2 & Method 3 & Method 4 & Method 5 & Method 6 \\ \hline
			Eq. &  (\ref{eq.tier1}) &  (\ref{eq.tier2}) &  (\ref{eq.of}) &  (\ref{eq.m}) &  (\ref{eq.m}) &  (\ref{eq.carbonm})\\
			Tier & 1 & 2 & 1 & 1 & 2 & 3 \\
			EF & IPCC & ISPRA & IPCC & IPCC & ISPRA & IPSRA\\
			\bottomrule
		\end{tabular}
	}
\end{table}

\begin{table}[h]
	\caption{Fuel parameters and emission factors used in the models.}
	\label{tab.fuelp}
	\centering 
	\resizebox{12cm}{!}{
		\begin{tabular}{lcccc}
			\toprule
			& Emission factor & Emission factor& Oxidation rate & Net calorific value \\
			& (tCO$_2$/TJ) & (tCO$_2$/MWh) & (unit) & (TJ/10$^3$ ton) \\
			\midrule
			IPCC  & & &\\
			Fossil coal &  94.60 & 0.34 & 0.92 & 20.91\\
			Derived gas &  107.07& 0.39 & 0.93 & 33.46 \\
			Natural gas & 56.10  & 0.20 & 0.99 & 38.93\\
			Fossil oil & 77.7    & 0.28 & 0.98 & 41.82\vspace{0.2cm}\\
			ISPRA & & &\\
			Fossil coal &  94.13 & 0.34 & 1 &\\
			Derived gas &  163.36$^a$ & 0.59$^a$ & 1 &\\
			Natural gas & 56.38 & 0.20 & 1 &\\
			Fossil oil & 76.59 &  0.28 & 1 & \\
			\bottomrule
		\end{tabular}
	}
	\begin{tablenotes}
		\item[]{\footnotesize $^a$It is computed as the average value of three ISPRA emission factors: oxygen steel mill gas, blast furnace gas, and coke oven gas. Conversion factor: 1 TJ is 277,7778 MWh. Data source: \cite{IPCC2006} and \cite{ISPRA}.}
	\end{tablenotes}
\end{table}
More specifically, we use hourly net electricity generation data (in MWh), for each energy source and for each zone of the Italian day-ahead electricity market, to compute the zonal hourly emissions from electricity generation for the period 2016-2023. This data was sourced from \cite{entsoe}.

The primary objective of this comparison is to quantify the differences among the different approaches and to investigate whether, and to what extent,  more complex methods yield more accurate estimates of the emission. It is important to note that, for some methods, a direct comparison may be unfeasible due to the lack of readily detailed available information. For example, we will restrict the comparison with the \cite{Beltrami2021a} method to the year 2018 and Northern Italy due to data availability limitations. Calculating CO$_2$ emissions using this method poses challenges (Tier 3), so we conduct a direct comparison with the results presented in \cite{Beltrami2021a}.

Although we compute CO$_2$ emissions at hourly frequency, the methods' comparison is made first considering monthly hourly average, where the average is done both month-by.month and on 30-day-moving windows, and then aggregating the estimates year-by-year. In addition, we work separately for each of the seven zones of the Italian electricity market (namely, North, Centre-North, Centre-South, South, Calabria, Sicily and Sardinia) to be able to highlight possible differences. In the following, we will focus first on the Northern zone, which is the most important and the most emissive zone, and then we will consider all other zones. This will allow us to assess both the differences between methodologies and the trends in CO$_2$ emissions in the various regions of the Italian electricity market, providing information on the effectiveness and accuracy of different estimation techniques.

\subsection{The North zone}
\begin{figure}[H]
\centering
\includegraphics[width=1\textwidth]{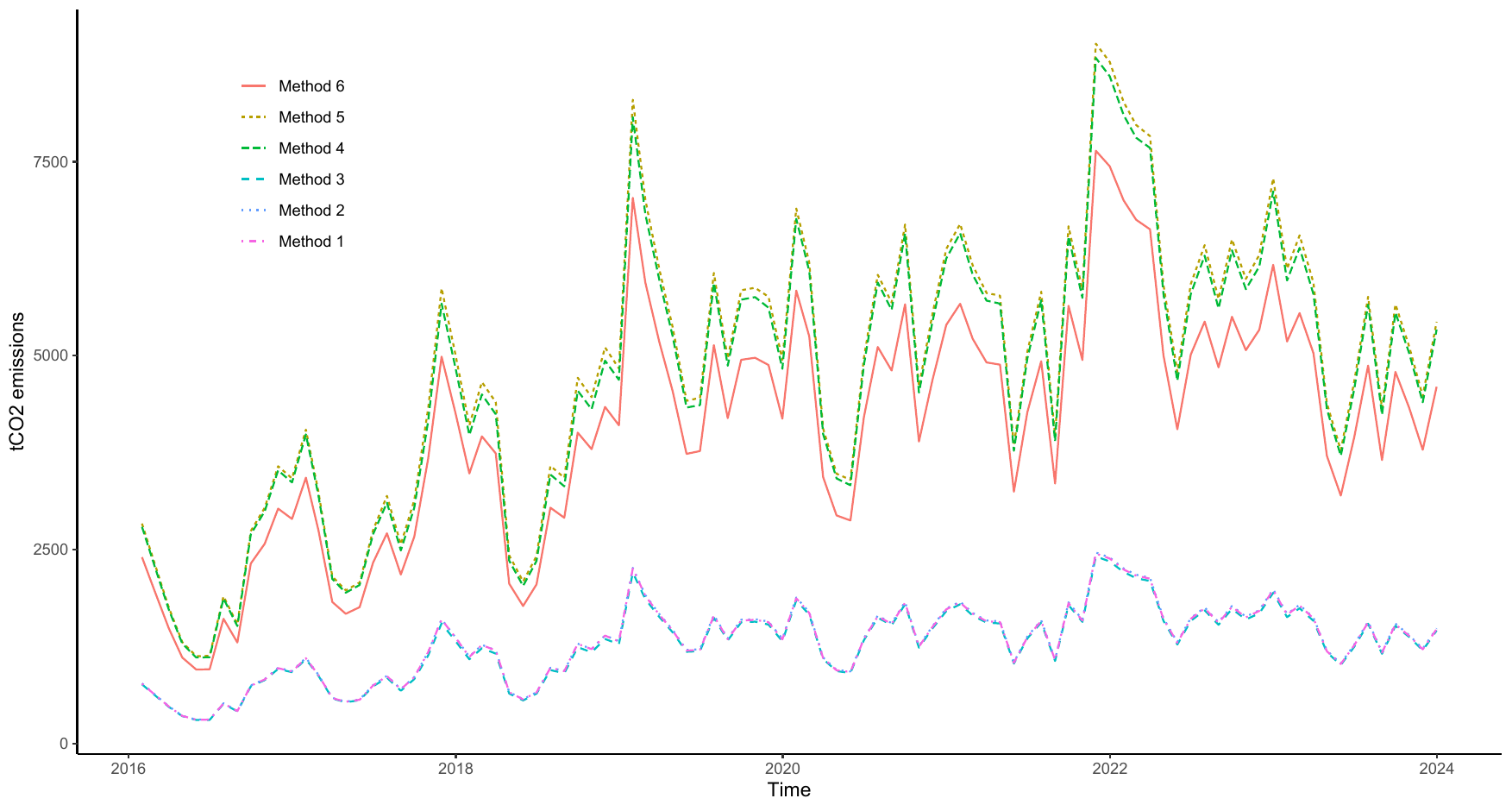}
\caption{Monthly mean of CO$_2$ estimates for the market zone North.}\label{fig.monthmean}	
\end{figure}

\begin{figure}[H]
\centering
\includegraphics[width=1\textwidth]{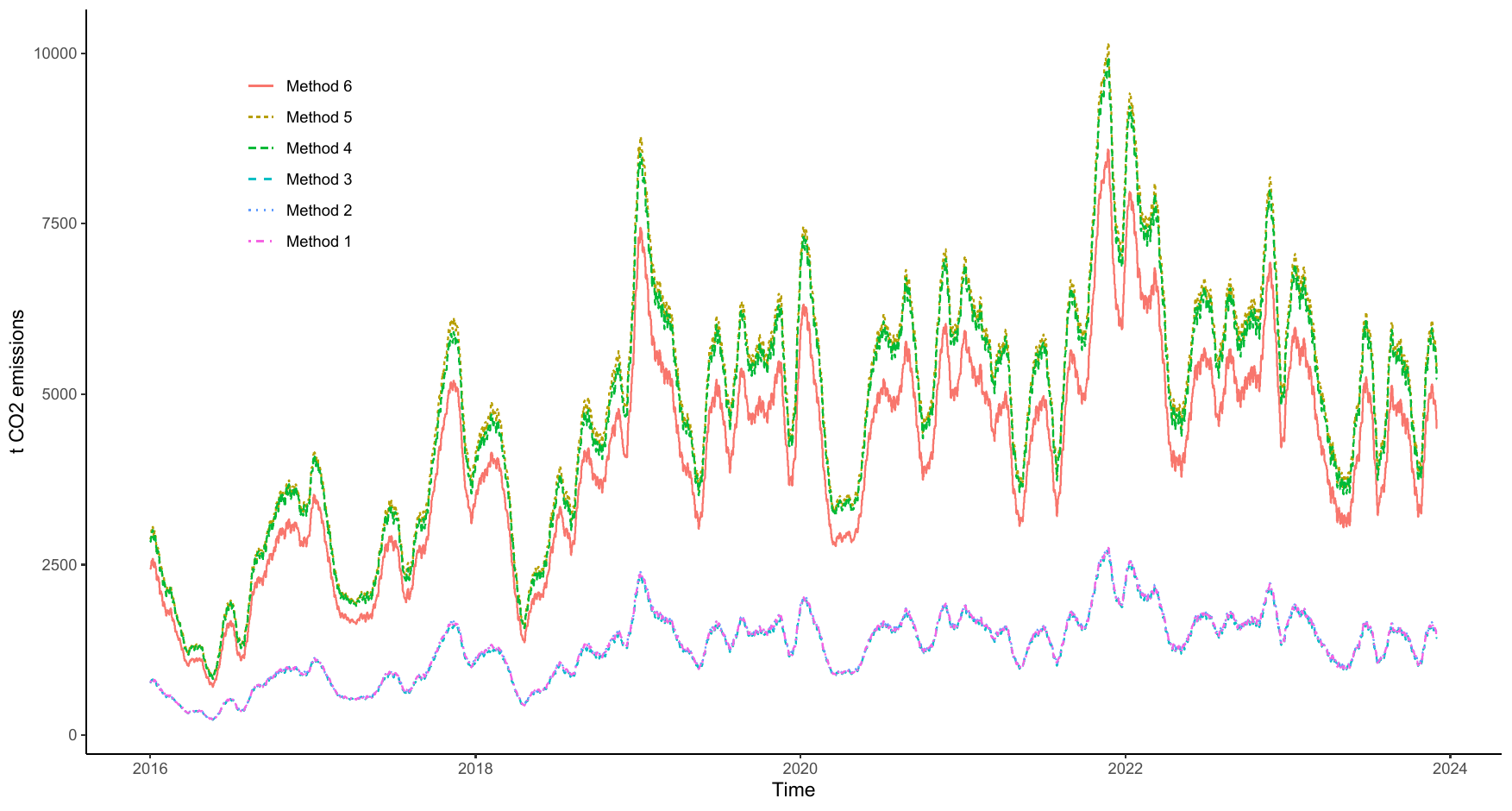}
\caption{Rolling mean (wd=720) of CO$_2$ estimates for the market zone North.}\label{fig.roolmean}	
\end{figure}

In this Section, we analyse the North zone, describing in detail our comparison. The next Section will contain results for all other zones.\\
Figure \ref{fig.monthmean} shows the monthly average of the CO$_2$ estimates for the North zone using the six methods in Table \ref{tab.methods}. All methods show an increasing trend in CO$_2$ emissions between 2016 and 2022, followed by a slowdown in 2023. However, this upward trend is less pronounced for methods 1-3, which suggests a more constant trend. The graph points out that different methods lead to substantially different level of estimated emissions. In particular, the methods can be grouped in two distinct clusters: methods 1,2 and 3 which lead to extremely similar  values of the emissions and differing among them, on average, less than 30 tCO$_2$ (see Figure \ref{fig.monthmean}). 
Also methods 4, 5 and 6 show quite similar patterns among them, but produce much higher values of the estimated emissions with respect to methods 1, 2 and 3. This suggests that they account more efficiently and accurately the actual emissions levels. In addition, while the emissions levels for methods 4 and 5 differ, on average, of around 100  tCO$_2$, method 6 underestimates the emissions with respect to methods 4 and 5, of around 700 tCO$_2$.\\
The average difference in estimated emission between method 5, which belongs to the high-level cluster, and Method 2, which belongs to the low-level cluster, is around 3500 tCO$_2$.\\
These figures clearly show that some calculation methods heavily underestimate emissions and, in the meantime, explain the importance of choosing the best possible method, coherently with the available information.  \\
It is interesting to note that the two groups are directly related to the tier-type classification. Rather, tiers cross the groups and, in particular, the two methods leading to the more exhaustive emissions estimates are methods 5 and 4, which belong, respectively, to Tier 2 and Tier 1. Method 6 is a Tier 3-type and belongs to the top group, but it seems underestimate emissions.
Finally, Tier 1 is represented by methods 4, 2 and 1 which, however, give quite different results as method 4 is in the top group while methods 2 and 1 are in the bottom group.\\
These results are further supported by the graph in Figure \ref{fig.roolmean}, which shows the analogous patterns when the average are computed on rolling windows of 720 (30 days).

\begin{figure}[H]
	\centering
	\includegraphics[width=1\textwidth]{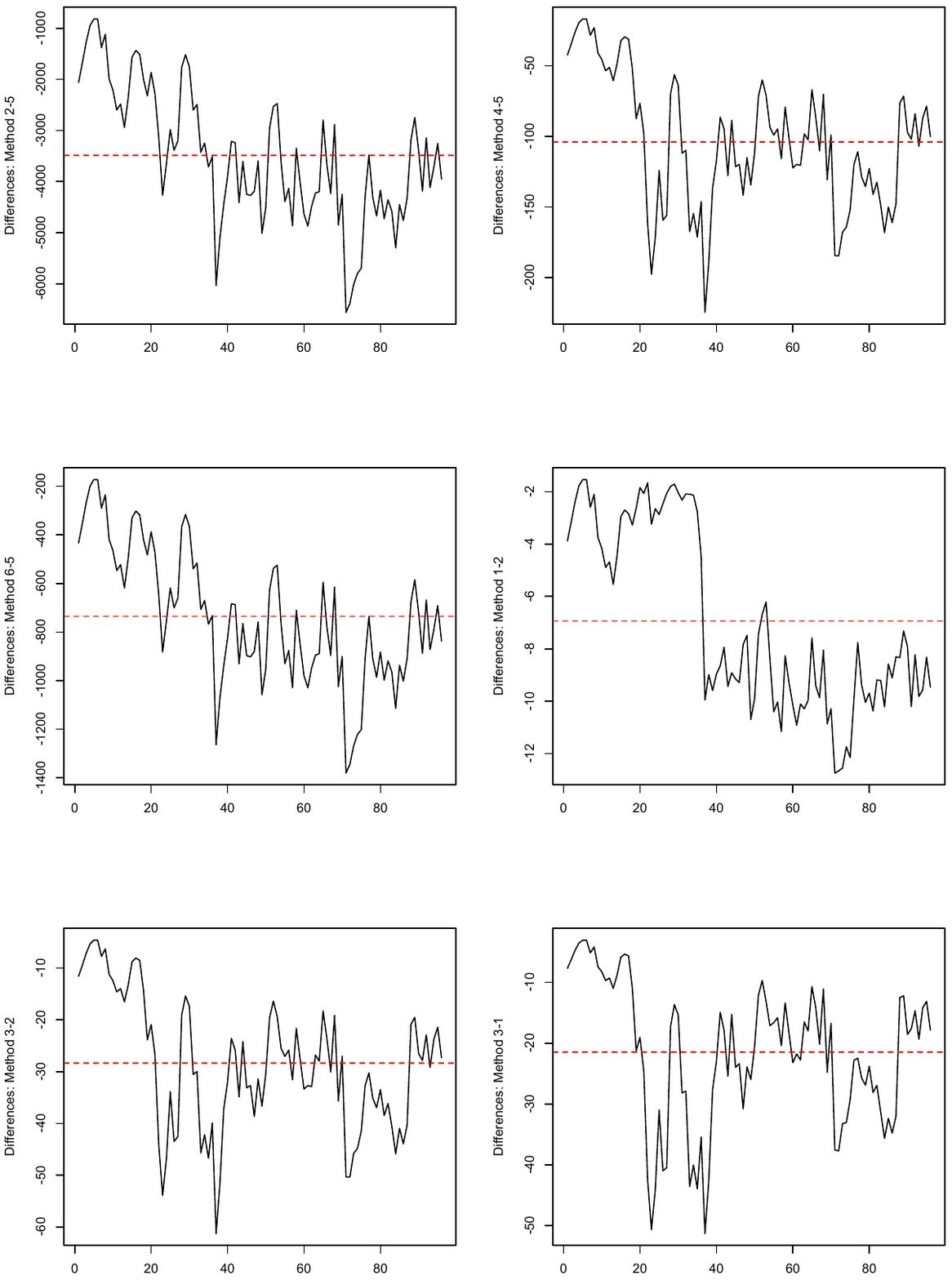}
	\caption{Monthly differences between some methods for the North (tCO$_2$).}\label{fig.diff_north}	
\end{figure}

Now we consider the issue of assessing the statistical significance of the mean difference (in tCO$_2$) between couples of methods.\\
Let be $d_{t,ij}=E_{t,i}-E_{t,j}$ the time series of the difference in the emission estimates computed using two different emission factor-based methods, say methods $i$ and $j$, and let be $\bar d_{ij}$ the sample mean.\\
Figure \ref{fig.diff_north} shows the series of these monthly differences between the methods belonging to the same group.\\
We want to test the hypothesis $H_0: E(d_{t,ij})=\mu$ against $H_1: E(d_{t,ij})\neq \mu$. As we have not independent data but a time series, we cannot use the classical t-test for the mean. Thus we resort to Diebold and Mariano (DM)-like test, which is not used to assess the predictive accuracy between models but, rather, the mean difference of two time series. Our test statistic is computed as follows:
\begin{equation} DM=\frac{\bar d-\mu}{\sqrt{(1/n) \sum_{i=-m}^{m}\gamma_i}
	} 
	\label{eq.DMtest} 
\end{equation}
\noindent where $\bar d$ is the observed mean difference between the two time series, $\mu$ is the mean under the null hypothesis and $\gamma_i$ is the covariance at lag $i$ of the series $d_{t,ij}$. According to \cite{Diebold1995}, $(1/n)\sum_{i=-m}^{m}\gamma_i$ is an estimator of the variance of $d_{t,ij}$.\\
We test null hypothesis of $\mu$ in order to find the values of $\mu$, and especially the minimum values $\mu_0$, leading to not reject $H_0$ at the $5\%$ significance level. In this context, $\mu_0$ represents our estimate of the average difference between each couple of methods. 

Table \ref{tab:dm_lower_triangular_approx} reports the results of the Diebold-Mariano test for different pairs of methods in the north zone, as indicated by their p-values in the range $\mu$. For the comparison between methods 4 and 5, the null hypothesis $H_0$ is not rejected at $\mu$ values ranging from -122 to -85. In the case of methods 6 and 4, $H_0$ is not rejected for $\mu$ values between 531 and 730, while for methods 6 and 5, the corresponding range is 621 to 849. Notably, for group 2, $H_0$ cannot be rejected for lower $\mu$ values compared to group 1, indicating smaller differences between the methods in the latter group.

\begin{table}[ht] 
\caption{Results of the Diebold-Mariano-like test for different method pairs in the North zone. Each cell contains the interval of values of \(\mu_0\) leading to not reject the null hypothesis that $E(d_{t,ij})=\mu_0$ at the \(5\%\) significance level.}\label{tab:dm_lower_triangular_approx} 
\centering 
\resizebox{15cm}{!}{ 
\begin{tabular}{ccccccc} 
\hline & Method 1 & Method 2 & Method 3 & Method 4 & Method 5 & Method 6 \\ \toprule Method 1 & - & & & & & \\ Method 2 & \([-9, -5]\) & - & & & & \\ Method 3 & \( [17, 26] \) & \( [23, 34] \) & - & & & \\ Method 4 & \( [-3917, -2870] \) & \( [-3909, -2865] \) & \( [-3942, -2889] \) & - & & \\ Method 5 & \( [-4037, -2959] \) & \( [-4028, -2954] \) & \( [-4061, -2978] \) & \( [-123, -85] \) & - & \\ Method 6 & \( [-3187, -2338] \) & \( [-3179, -2333] \) & \( [-3212, -2357] \) & \( [531, 731] \) & \( [621, 849] \) & - \\ \bottomrule 
\end{tabular} 
} 
\end{table}

As a further investigation, we consider estimated emissions aggregated at yearly frequency, i.e. the total yearly emissions.\\
Table \ref{tab.results} lists yearly CO$_2$ estimated emissions as well as the average yearly emission factors (AEF), for the market zone North. The AEF indicates the proportion of emissions produced to the amount of electricity generated during a specific year. 
Although the year with the largest electricity generation was 2019, all methods agree that the peak of CO$_2$ emissions was in 2022. For this year, the maximum level of  CO$_2$ emissions is around 57.5 millions of tCO$_2$/MWh and is obtained using method 5. As concerns the other methods, method 4 gives very similar results, underestimating the emissions of around $2\%$, method 6 underestimates emission of almost $11\%$, while methods 1, 2 and 3 lead to underestimate the emissions of around $73\%$. Again, these figures tell us that the methods are not equivalent. In addition, it seems that the adjustment for the oxidation rate (method 3) does not provide a significant improvement, at least in the current data.
Even if methods 4-5 provide quite similar estimates, Tier 2 methods tend to produce higher CO$_2$ estimates than Tier 1 methods because they use country-specific emission factors. 
Finally, it should be noted that for 2018, methods 4 and 5 produce estimates not so different from those of \cite{Beltrami2021b}, which employ the Tier 3 approach outlined in Eq. (\ref{eq.beltrami}). More specifically, CO$_2$ emissions and AEF calculated with (Eq.\ref{eq.beltrami}) for 2018 in northern Italy are 36,625,956 (tCO$_2$) and 0.2840 (tCO$_2$/MWh), respectively. As a result, there is a $8\%$ difference in CO$_2$ emissions and a $9\%$ difference in AEF between Method 5 and Eq.\ref{eq.beltrami}. There is then a small difference in electricity generation (2018), which is equivalent to 128,956,528 (MWh) for \cite{Beltrami2021b}.

\begin{table}[H]
\caption{CO$_2$ estimates for the market zone North grouped by IPCC tiers.}
\label{tab.results}
\centering 
\resizebox{15cm}{!}{
		\begin{tabular}{lccccccccc}
		\toprule
              Method &\multicolumn{7}{c}{Year}\\
              & 2016 & 2017 & 2018 & 2019 & 2020 & 2021 & 2022 & 2023\\
              \midrule
              &\multicolumn{7}{c}{\text{Tier 1}}\\
              Method 1\\
              $E_t$ & 5,292,676 & 8,004,418 & 9,167,253 & 13,669,145 & 12,676,531 & 14,517,971 & 15,606,947 & 12,290,343 \\ 
              AEF & 0.0441 & 0.0635 & 0.0701 & 0.1044 & 0.1014 & 0.1117 & 0.1277 & 0.1067\vspace{0.2cm}\\

              Method 3 \\
              $E_t$ & 5,239,749 & 7,821,361 & 8,891,957 & 13,436,191 & 12,525,581 & 14,334,421 & 15,359,359 & 12,116,383\\ 
              AEF & 0.0437 & 0.0620 & 0.0680 & 0.1026 & 0.1001 & 0.1103 & 0.1257 &0.1052\vspace{0.2cm}\\

              Method 4 \\
              $E_t$ & 19,212,413 & 28,678,324 & 32,603,844 & 49,266,034 & 45,927,131 & 52,559,544 & 56,317,649 & 44,426,737\\ 
              AEF & 0.1603 & 0.2273 & 0.2494 & 0.3763 & 0.3672 & 0.4043 & 0.4608 & 0.3856\vspace{0.6cm}\\
              
              &\multicolumn{7}{c}{\text{Tier 2}}\\
              Method 2 \\
              $E_t$ & 5,319,406 & 8,030,519 & 9,188,323 & 13,746,683 & 12,756,011 & 14,607,524 & 15,696,088 & 12,367,095\\ 
              AEF & 0.0444 & 0.0637 & 0.0703 & 0.1050 & 0.1020 & 0.1124 & 0.1284 & 0.1073\vspace{0.2cm}\\

              Method 5 \\
              $E_t$ & 19,504,490 & 29,445,235 & 33,690,519 & 50,404,506 & 46,772,042 & 53,560,921 & 57,552,323 & 45,346,014\\ 
              AEF & 0.1627 & 0.2334 & 0.2577 & 0.3850 &  0.3740 & 0.4120 & 0.4709 & 0.3936\vspace{0.6cm}\\

              &\multicolumn{7}{c}{\text{Tier 3}}\\
              Method 6 \\
              $E_t$ & 17,373,608 & 26,275,107 & 30,092,200 & 44,870,000 & 41,611,671 & 47,656,338 & 51,230,984 & 40,343,980\\
              AEF & 0.1449 & 0.2083 & 0.2302 & 0.3427 & 0.3327 & 0.3666 & 0.4192 & 0.3501\vspace{0.6cm}\\

              $G_t$ & 119,882,362 & 126,143,605 & 130,747,043 & 130,922,842 & 125,075,027 & 129,996,169 & 122,222,342 & 115,208,515\\
		\bottomrule
		\end{tabular}
  }
  \begin{tablenotes}
\item[]{\footnotesize \textit{Notes}. The term $E_t$ denotes the annual CO$_2$ emissions (tCO$_2$), while $G_t$ (MWh) denotes the yearly net electricity generation. AEF (tCO$_2$/MWh) is the average emission factor.}
\end{tablenotes}
\end{table}

\subsection{The other zones}
\begin{figure}[H]
	\centering
	\includegraphics[width=0.8\textwidth]{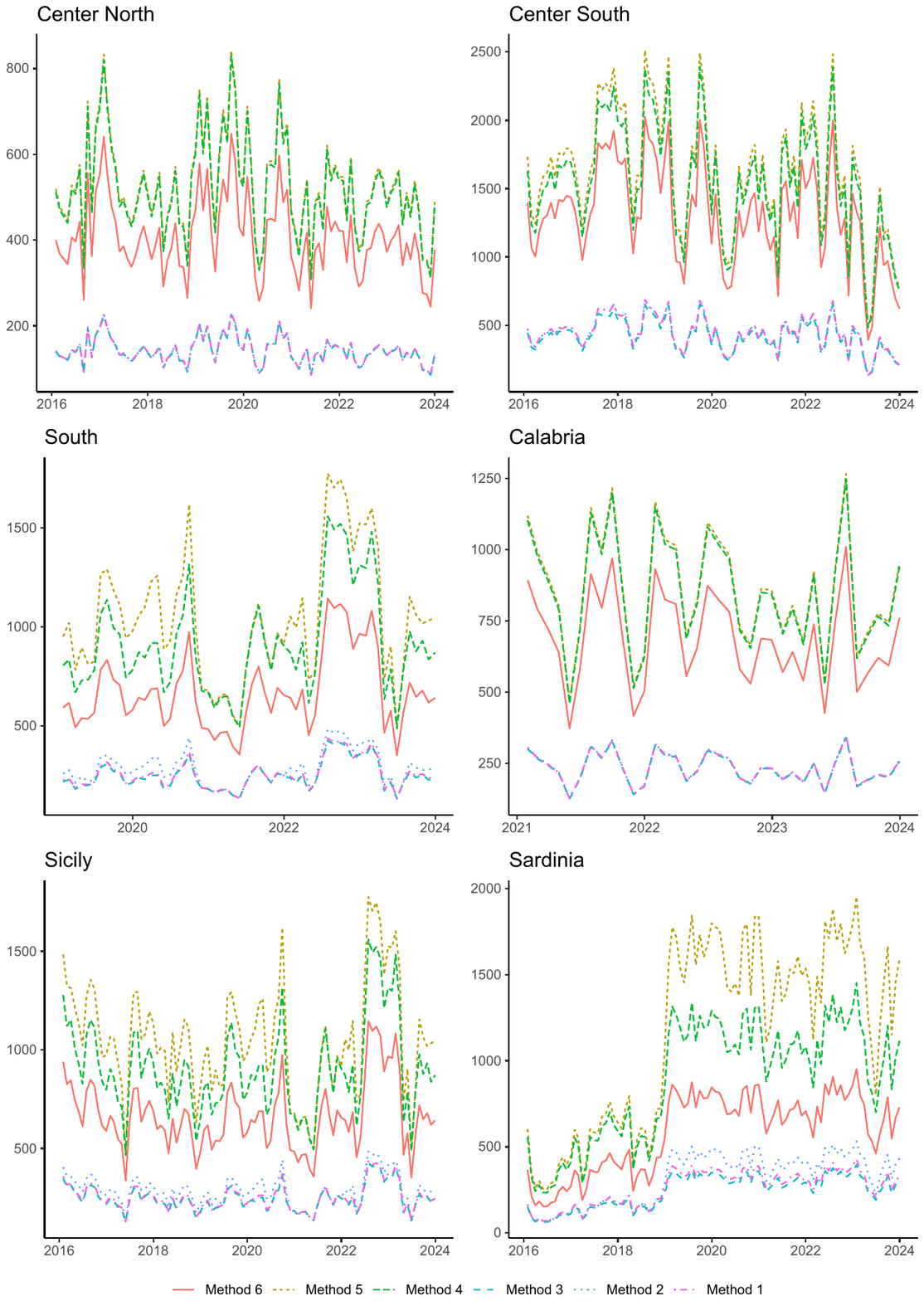}
	\caption{Monthly average zonal carbon emissions (tCO$_2$) in Italy: comparison of estimation methods. Data are unavailable for the South zone from 2016-2018 and for Calabria from 2016-2020.}
	\label{fig.monthlym_zones}
\end{figure}

The analyses conducted for the north zone were also carried out for the other zones. The zonal analysis is interesting because it is known that they have different features and different fuel mixes, especially Sardinia and Sicily, which are islands.\\  	
Figure \ref{fig.monthlym_zones} shows the estimated monthly hourly average zonal CO$_2$ emissions in Italy from 2016 to 2023, within the Center-North, Center-South, South, Calabria, Sicily and Sardinia zones. 

First of all, it is worth noting that the global emission levels in the North zone are much higher than in the other zones. This is partially connected with the higher production level, but non only, as it is clear in the case of Sardinia.\\
With respect to the long term behaviour we can distinguish three groups: Center-North and Center-South, which show a slight downward trend, at least in the last years; South, Calabria and Sicily, which do not show a clear pattern; and Sardinia, which seems to show an increase in the emission rate during the last years.

\begin{figure}[H]
	\centering
	\includegraphics[width=1\textwidth]{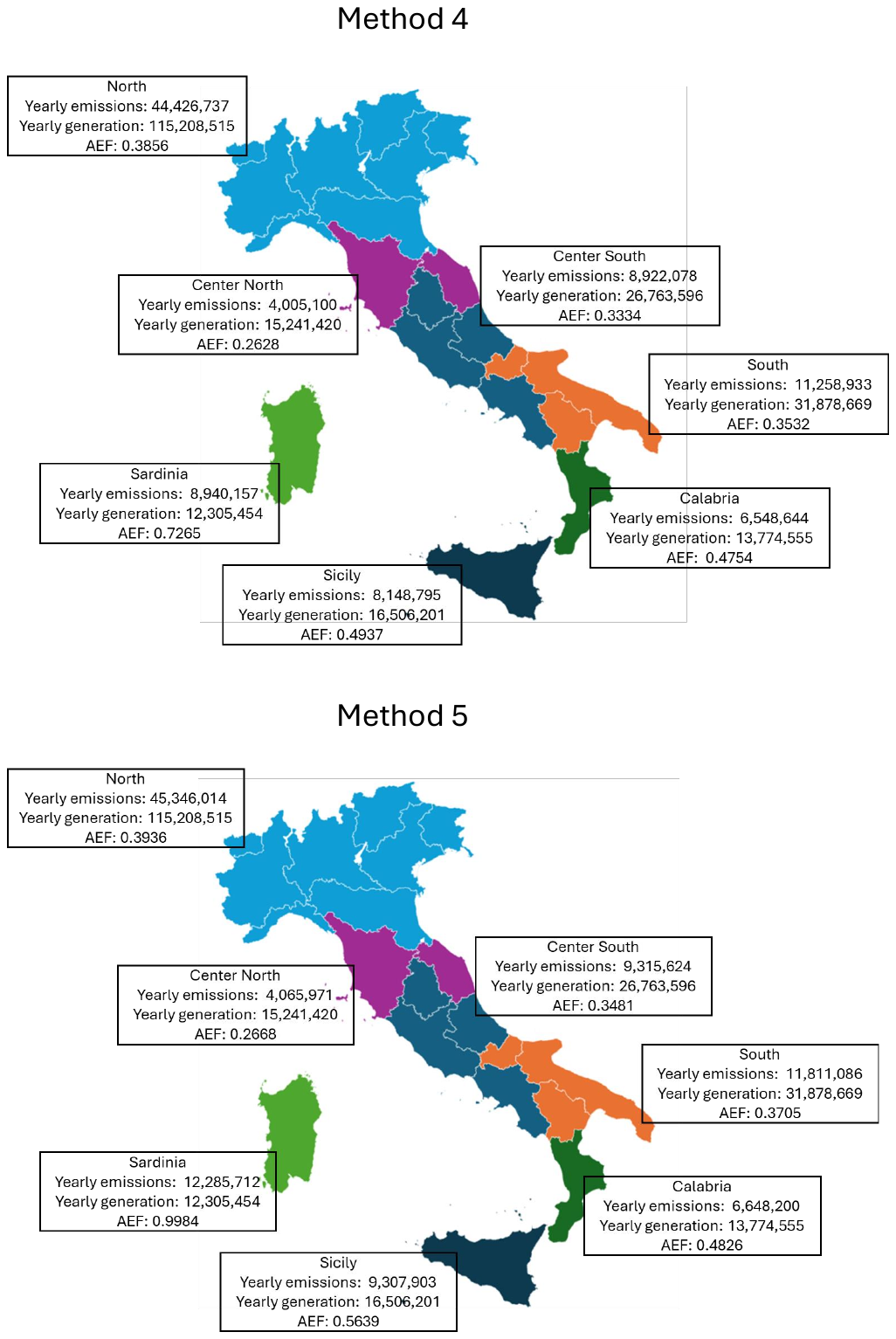}
	\caption{Zonal configuration of the Italian electricity market. Data refer to year 2023. Values of generation are in MWh, values of carbon emissions are in tCO2, and AEF values in tCO$_2$/MWh.}
	\label{key_ind}
\end{figure}

With respect the performance of the different methods, Center-North and Center-South zones appear to behave similarly to the north zone with methods 1,2 and 3 producing a first cluster and methods 4, 5 and 6 leading to a second cluster. For the remaining zones, methods tend to differentiate among them, even if methods 1, 2 and 3 continue to produce much lower levels of estimated emissions. For South and Calabria, three distinct clusters emerge due to method 6 yielding significantly lower estimates than methods 4 and 5. The islands, Sicily, and Sardinia, exhibit specific patterns that set them apart from the other areas. In this zones, the contrast between methods that use default emission factors and those that use country-specific emission factors is more pronounced because it is well known that the generation of electricity from fossil coal and derived gas is higher compared to other market zones (Table \ref{tab.fuelp}). Although country-specific emission factors generally align with IPCC values, derived gases are an exception: the default emission factor is 0.39 (tCO$_2$/MWh), while the country-specific emission factor is 0.59 (tCO$_2$/MWh). Therefore, the adoption of one method instead of another for Sicily and Sardinia is extremely relevant. For example, Figure \ref{key_ind} shows the zonal configuration of the Italian electricity market, with data from the recent year 2023. 

The estimated AEF by methods 4-5 change significantly for Sicily and Sardinia. According to method 4 the AEF is 0.5939 (tCO$_2$/MWh) and 0.9984 (tCO$_2$/MWh) for Sicily and Sardinia, respectively. In contrast, method 5 estimates the AEF at 0.4937 (tCO$_2$/MWh) and 0.7265 (tCO$_2$/MWh).

The results of these analyses suggest several interesting considerations. Method 3, which incorporates the oxidation factor, does not particularly affect the emission estimates. Methods 2 and 5, which are based on specific emission factors, do not have a relevant impact on the emission estimates for the North, Center North, Center South, South, and Calabria. In contrast, specific emission factors do significantly influence the emission estimates for Sicily and Sardinia, as these zones have a higher proportion of derived gases, and the specific emission factors used by Methods 2 and 5 differ from the IPCC standards. Moreover, Methods 4 and 5, which account for the molecular weight ratio of carbon dioxide to carbon, yield higher and more consistent estimates compared to Methods 1, 2, and 3. Interestingly, Method 6, which follows the Tier 3 approach, does not necessarily produce more accurate estimates. Additionally, the Tier 3 method proposed by Beltrami (2021) provides estimates comparable to those of Methods 4 and 5 (Tier 1-2), while avoiding complex calculations and the need for specific plant-level data, the source of which remains unclear in Beltrami's work.

\section{Policy implications}\label{sec.policy}
For a long time, carbon emissions represented negative externalities with no markets and no monetary costs associated, but the increasing concern about environment led to an evolution in policy making, and the creation of markets changed their role.

In the development of national and supranational policies, it has become increasingly important to account for emissions related to different goods' categories, also considering their entire life cycle (from production to disposal). The same approach is applied to processes in all relevant sectors \citep{RAHN2024142195,WANG2019471,COMODI2016944}. 

As mentioned above, one of the most relevant experience based on emissions and their trading in Europe is the European Union Emissions Trading System (EU ETS). The ETS was launched in 2005 with the aim of making polluters pay for the costs of carbon emissions: the system currently includes electricity and heat generation, industrial manufacturing, and aviation sectors, plus the maritime transport sector that entered the system in 2024. Thanks to this system, polluting industrial sectors are required to monitor and report their emissions and keep them below the limit set by the allowances system. Allowances can be traded to keep the company within the threshold set by the cap system \citep{ETSCons}.

The European Commission claims that, in addition to the effects in terms of reduction in direct carbon emissions, the system contributed to the energy transition process with more than 175 billion euros \citep{eu_ets}, which are invested in the energy transition by the EU through the Innovation Fund \citep{innovation_fund} and the Modernisation Fund \citep{modernisation_fund}. 

The presence of ETS also influences corporate decision making, with impacts on other sectors \citep{LEE2024142188}. In this context, monitoring and reporting emissions became more relevant with the introduction of sustainability criteria for ESG (environmental, social, and governance) to evaluate companies.

Consistent with the evidence highlighted in the scientific literature and in accordance with international agreements and policies from different regions, concrete actions and several projects have been evaluated with respect to CO$_2$ reduction. Where this decrease is not possible, carbon offset actions \citep{carbon_offsets} are encouraged or required. The European funding programme INTERREG-MED recently implemented the tool ZeroCO$_2$MED to include carbon offset directly in the project design and in the evaluation phase: this means that to access the fund it is necessary to first calculate its impact in terms of emissions and choose how to compensate them \citep{carbon_footprint_meeting}.
Our work highlighted how the application of different calculation methodologies can lead to significant discrepancies in the final results. In a market environment such as those where the ETS system is present or, prospectively, in broader markets that consider environmental criteria as economically relevant, this can have repercussions on the competitiveness of companies, potentially distorting the market. What is necessary is that, at the policy level, for each sector and market segment, it should also be determined which methodology to apply, to avoid creating disparities and distortions.

\section{Conclusions}\label{sec.conclusions}

In this study, several methods to calculate CO$_2$ emissions from electricity generation have been evaluated. The analyses showed significant differences in accuracy depending on the method used. Methods that incorporate country-specific and technology-specific data, i.e. higher-tier methods, tend to provide more accurate CO$_2$ estimates compared to basic methods, i.e. Tier 1 methods, that use default emission factors. However, despite this theoretical result, our practical application showed that, if applied in the correct context, also TIER 1 methods (method 4 for the northern area) can perform well.

Results suggest that effective estimation of CO$_2$ emissions can be performed using methods 4-5. However, it is crucial to assess whether the default emission factors align with those of a specific country or zone. For Italy, the emission factors are largely consistent with the IPCC values, except for derived gases. In this case, the use of IPCC or ISPRA factors does not significantly affect the results, nor does the inclusion of the oxidation coefficient which is close to 1. Alternatively, method 6 provides a reliable estimate by adjusting default emission factors with country-specific baseline emissions (indirectly integrating country-specific information).

The Italian zonal CO$_2$ emissions analysis underscores the importance of using country-specific emission factors and also zone-specific factors if present, especially in areas such as Sicily and Sardinia, where fossil fuels such as coal and derived gases play a substantial role in electricity generation.
In these zones, the observed differences between the default and country-specific emission factors are pronounced.\\
This kind of discrepancy mirrors the practice within the European Union, where many countries apply country-specific emission factors to improve the accuracy of greenhouse gas inventories, in line with United Nations Framework Convention on Climate Changes (UNFCCC) recommendations to reduce uncertainties \citep{LI2024122681}. The comparative analysis of methods 4 and 5 in Sicily and Sardinia further demonstrates that selecting appropriate emission factors is crucial for reliable emission estimation, as the resulting AEF values vary significantly between methods. This approach ensures a better alignment with the regional fuel characteristics, offering a more precise reflection of the environmental impact associated with fossil fuel use in the Italian electricity market zones. Therefore this result does not only occur between European countries but also between areas within them.\\
The comparison shown in this work is relevant if one considers the impact that emission calculation can have on company balance sheets and national performances, especially in relation to policy objectives.
Given that legislative requirements and company results are based on the calculation of emissions and considering the monetary value associated with them, future research should extend these evaluations to other European or international electricity markets and to other sectors to understand how to improve the methods when applied.

In general, adopting detailed and specific approaches will improve the understanding and management of emissions in the evolving global energy landscape, but operations still matter.

\paragraph{Declaration of interests} The authors report that financial support was provided by European Union - NextGenerationEU, Mission 4, Component 2, in the framework of the GRINS - Growing Resilient, INclusive and Sustainable project (GRINS PE00000018 - CUP C93C22005270001).

\paragraph{Acknowledgments} This study was funded by the European Union - NextGenerationEU, Mission 4, Component 2, in the framework of the GRINS - Growing Resilient, INclusive and Sustainable project (GRINS PE00000018 - CUP C93C22005270001). The views and opinions expressed are solely those of the authors and do not necessarily reflect those of the European Union, nor can the European Union be held responsible for them.

\bibliographystyle{apalike}
\bibliography{References.bib}

\end{document}